\documentclass[%
 reprint,
 amsmath,amssymb,
 prd,
]{revtex4-1}
\usepackage{url,amssymb,amsmath,rotating,color,units,wasysym,epsfig,multirow,epstopdf}
\usepackage[colorlinks,urlcolor=blue,citecolor=blue,linkcolor=blue]{hyperref}

\usepackage{graphicx}
\usepackage{dcolumn}
\usepackage[normalem]{ulem}
\usepackage{bm}
\usepackage[mathlines]{lineno}

\newcommand{\gwtcspeedup}{$\sim 10\times$ }

\newcommand{\cpu}{Intel Gold 6140}
\newcommand{\gpu}{NVIDIA P100}

\newcommand{\SPA}{School of Physics and Astronomy, Monash University, Vic 3800, Australia}
\newcommand{\OzGravMonash}{OzGrav: The ARC Centre of Excellence for Gravitational Wave Discovery, Clayton VIC 3800, Australia}

\begin{document}


\title{Parallelized Inference for Gravitational-Wave Astronomy}

\author{Colm Talbot}
\thanks{colm.talbot@monash.edu}
\affiliation{\SPA}
\affiliation{\OzGravMonash}
\author{Rory Smith}
\affiliation{\SPA}
\affiliation{\OzGravMonash}
\author{Eric Thrane}
\affiliation{\SPA}
\affiliation{\OzGravMonash}
\author{Gregory B. Poole}
\affiliation{
Astronomy Data and Computing Services (ADACS); the Centre for Astrophysics \& Supercomputing, Swinburne University of Technology, P.O. Box 218, Hawthorn, VIC 3122, Australia}

\date{\today}

\begin{abstract}
Bayesian inference is the workhorse of gravitational-wave astronomy, for example, determining the mass and spins of merging black holes, revealing the neutron star equation of state, and unveiling the population properties of compact binaries.
The science enabled by these inferences comes with a computational cost that can limit the questions we are able to answer.
This cost is expected to grow.
As detectors improve, the detection rate will go up, allowing less time to analyze each event.
Improvement in low-frequency sensitivity will yield longer signals, increasing the number of computations per event.
The growing number of entries in the transient catalog will drive up the cost of population studies.
While Bayesian inference calculations are not entirely parallelizable, key components are embarrassingly parallel: calculating the gravitational waveform and evaluating the likelihood function.
Graphical processor units (GPUs) are adept at such parallel calculations.
We report on progress porting gravitational-wave inference calculations to GPUs.
Using a single code---which takes advantage of GPU architecture if it is available---we compare computation times using modern GPUs (\gpu) and CPUs (\cpu).
We demonstrate speed-ups of $\sim50\times$ for compact binary coalescence gravitational waveform generation and likelihood evaluation, and more than $100\times$ for population inference within the lifetime of current detectors.
Further improvement is likely with continued development.
Our python-based code is publicly available and can be used without familiarity with the parallel computing platform, CUDA.
\end{abstract}

\pacs{Valid PACS appear here}
\maketitle

\section{\label{sec:motivation}Introduction}
In the first two observing runs of Advanced LIGO/Virgo, ten binary black hole mergers were detected along with one binary neutron star inspiral~\cite{GWTC1}.
These observations allowed us to measure the Hubble parameter~\cite{GW170817Hubble}, to study matter at extreme densities~\cite{GW170817EoS}, and to probe the underlying distribution of black holes in merging binaries~\cite{O2Pop}.
Within the lifetime of advanced detectors, we conservatively estimate that hundreds of such observations will be made given inferred merger rates and projected sensitivity~\cite{Abbott2016}.

Compact binary coalescences are analyzed with Bayesian inference (see, e.g.~\cite{gelman2013bayesian} for a general introduction or~\cite{Thrane2019} for applications to gravitational waves.).
We distinguish between two kinds.
We refer to inferring the properties (e.g., the masses, spins, and location) of individual binaries as {\em single-event inference}.
Hierarchical Bayesian inference is then used to infer the ensemble properties (e.g. the shape of the binary black hole mass distribution) of the observed binaries in {\em population inference}.
These are typically performed with stochastic sampling algorithms such as Markov Chain Monte Carlo (MCMC)~\cite{Metropolis1953, Hastings1970} or Nested Sampling~\cite{Skilling04}.
These algorithms generate samples from the posterior distribution and possibly a Bayesian evidence which can be used for model selection.

Both single-event inference and population inference require the computation of likelihood functions consisting of many independent operations.
For single-event inference, the number of operations per likelihood evaluation is determined by the length of the signals being analyzed, see, Eq.~\ref{eq:cbclike}.
As the low-frequency sensitivity of detectors increases, binaries will spend longer in our sensitive frequency range, leading to fast-increasing computational demands.
For population inference, the number of operations required per likelihood calculation is proportional to the number of binaries in the population, see, Eq.~\ref{eq:poplike}.
The growing number of observations and the growing duration of the longest signals in the catalog require improved speed for inference algorithms.

Most previous methods for accelerating inference for compact binary coalescences have sped up calculations by reducing the amount of data required to represent the gravitational-wave signal, thereby reducing the number of operations required to evaluate the likelihood, e.g., reduced order methods~\cite{Purrer2014, Canizares2015, Smith2016}, multi-banding~\cite{Vinciguerra2017}, and relative binning~\cite{Zackay2018}.
Another approach, which we investigate here, is to parallelize the most time-consuming calculations in the likelihood evaluation.
While it is difficult to parallelize the actual sampling algorithm, there are embarrassingly parallel calculations within the likelihoods.
In this paper, we explore how astrophysical inference can be accelerated by executing parallelizable calculations on graphical processor units.

While we focus here on inference using stochastic samplers, e.g.,~\cite{Veitch2015, Biwer2019, Ashton2019}, it bears mentioning that there are alternative inference schemes, which also face computational challenges, e.g., iterative fitting~\cite{Pankow2015, Lange2018} is one such alternative.
This method evaluates far fewer gravitational waveforms, which can significantly reduce computation times for inference when the waveform evaluation is very time-consuming.
Recently, it has been shown that this algorithm can also be significantly accelerated by parallelization~\cite{Wysocki2019}.

Modern computation is mostly performed on either a central processing unit (CPU) or a graphics processing unit (GPU).
CPUs consist of a relatively small number of cores optimized to perform all the tasks necessary for a computer in serial.
GPUs, on the other hand, consist of hundreds or thousands of cores, enabling evaluation of many numerical operations simultaneously.
This makes them ideal for embarrassingly parallel operations such as manipulating large arrays of numbers.
Low-level GPU programming is carried out using the parallel computing platform, CUDA.
In this work, we take advantage of the library, \textsc{cupy}~\cite{Okuta2017}.

We present python packages with parallelized versions of the likelihoods necessary for performing both single-event and population inference.
The code is designed to run on either a CPU or a GPU, depending on the available hardware.
Our GPU-compatible code for single-event inference is available at \url{github.com/colmtalbot/gpucbc}, our CUDA compatible version of \textsc{IMRPhenomPv2} at \url{https://github.com/ADACS-Australia/ADACS-SS18A-RSmith}.
Our GPU population inference code \textsc{GWPopulation} is available from the python package manager \textsc{pypi} and from git at \url{github.com/colmtalbot/gwpopulation}.

Using our GPU-accelerated code, we investigate the speed-up achieved carrying out gravitational-wave inference calculations using GPUs versus traditional CPUs.
We carry out a benchmarking study in which we compare inference code using GPUs to identical code running on CPUs.
We compare the runtimes for various tasks including: (i) evaluating a gravitational waveform, (ii) evaluating the single-event likelihood, and (iii) evaluating population likelihood.
To carry out this comparison, we use \gpu~GPUs and \cpu~CPUs available on the OzStar supercomputing cluster.

In Section~\ref{sec:wf} we describe methods for parallelizing the evaluation of gravitational waveforms.
In section~\ref{sec:cbc} we use these parallelized waveforms to consider the possible acceleration for single-event inference.
In section~\ref{sec:popinf}, we investigate the possible acceleration from parallelizing the population inference likelihood.
Finally, we provide a summary of our findings and future work in section~\ref{sec:discussion}.

\section{\label{sec:wf}Waveform Acceleration}

A key ingredient for single-event inference is a model for the waveform, $\tilde{h}(f, \theta)$.
Here, $\tilde{h}$ is the discrete Fourier transform of the gravitational-wave strain time series, $f$ is frequency, and $\theta$ is the set of parameters (typically 15-17), which determine the waveform, e.g., the masses, spins, and orientation of the binary.
This theoretical waveform is compared with the data every time the likelihood function is evaluated.
Many commonly used waveform models can be directly evaluated in the frequency domain and the value at each frequency can be evaluated independently (e.g.,~\cite{Ajith2011, Hannam2014, Khan2016, Khan2018}).
This makes the waveform model evaluation embarrassingly parallel.

We design two different codes for performing waveform generation on a GPU.
First, we implement a GPU version of a simple, post-Newtonian inspiral-only waveform \textsc{TaylorF2}~\cite{Ajith2011} directly in python using \textsc{cupy}.
Secondly, we convert the C code for a phenomenological waveform \textsc{IMRPhenomPv2}~\cite{Hannam2014} into CUDA\footnote{Specifically, we rewrite in CUDA the function lalsimulation.SimIMRPhenomPFrequencySequence branched from LALSuite at SHA:8cbd1b7187}.
Both of these methods are then compared to the time required to evaluate the same waveform implemented in C within \textsc{LALSuite}~\cite{LALSuite}.

Technically, the CUDA version of \textsc{IMRPhenomPv2} does not adhere to our philosophy of keeping the GPU programming entirely ``under the hood,'', however, writing custom CUDA kernels allows increased optimization of the GPU code.
We consider the effect of pre-allocating a reusable memory ``buffer'' on the GPU.
During parameter estimation waveforms of the same length will be generated many times as a waveform evaluation is required for every likelihood evaluation.
Since the waveform always has the same size a predefined spot in memory can be reused for every waveform.

\begin{figure}[!t]
    \centering
    \includegraphics[width=\columnwidth]{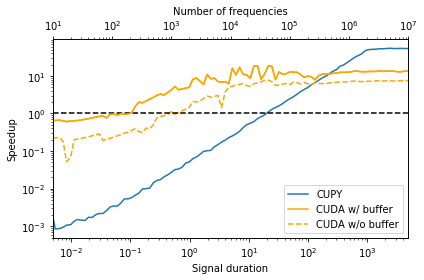}
    \caption{
    Speedup comparing our GPU accelerated waveforms with the equivalent versions in LALSuite as a function of the number of frequencies.
    The corresponding signal duration is shown for comparison assuming a maximum frequency of 2048 Hz.
    The blue curve shows the comparison of our \textsc{cupy} implementation of \textsc{TaylorF2} with the version available in lalsuite.
    The GPU version is slower for shorter waveform durations, this is likely because of overheads in memory allocation in \textsc{cupy}.
    For binary neutron star inspirals the signal duration from $\unit[20]{Hz}$ is $\sim\unit[100]{s}$, at which signal durations, we see a speedup of $\sim 10\times$.
    For even longer signals we find an acceleration of up to $80\times$.
    The orange curves show a comparison of our \textsc{CUDA} implementation of IMRPhenomPv2, the solid curve reuses a pre-allocated memory buffer while the dashed line does not.
    We see that the memory buffer makes the GPU waveform faster for shorter signals.
    }
    \label{fig:phenompspeedup}
\end{figure}

In Fig.~\ref{fig:phenompspeedup}, we plot the speed-up (defined as the CPU computation time divided by the GPU computation time) for both of our waveforms.
The dashed line indicates no speedup.
In blue we show the comparison of our \textsc{TaylorF2} waveform using \textsc{cupy} with the C version available in lalsuite~\cite{LALSuite}.
We find that, for signals longer than $\sim\unit[10]{s}$, we achieve faster waveform evaluation.
The speed-up scales roughly linearly with signal duration so that the waveforms of duration $\unit[100]{s}$ are sped up by a factor of $\approx10$.
For signals longer than $\sim\unit[1000]{s}$ the GPU queue saturates and the rate of increase of the speedup decreases.

In orange, we plot the speedup for our CUDA implementation of IMRPhenomPv2 compared to the C implementation of the same waveform model in lalsuite.
The solid curve includes the use of a pre-allocated memory buffer, while the dashed curve does not.
We find that pre-allocating memory buffer increases the performance when the number of frequencies is $\lesssim 10^5$ and leads to accelerations for all waveforms when the number of frequency bins is larger than 100.
This suggests that the performance of our \textsc{cupy} waveform could be similarly improved for short signal durations using more sophisticated waveform allocation.
%

\section{\label{sec:cbc}Single-event Likelihood acceleration}
\begin{figure}[!t]
    \centering
    \includegraphics[width=\columnwidth]{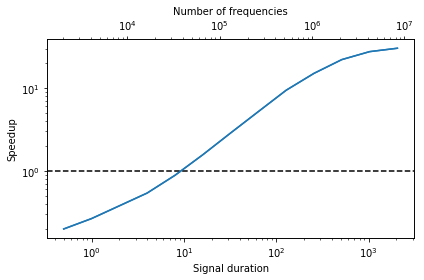}
    \caption{
    Speedup of the compact binary coalescence likelihood as a function of the number of frequencies with and without a GPU using our accelerated TaylorF2.
    The corresponding signal duration is shown for comparison assuming a maximum frequency of 2048 Hz.
    We see similar performance as for the waveform generation.
    We find a speedup of an order of magnitude for 128s signals.
    }
    \label{fig:cbclikespeedup}
\end{figure}

The standard likelihood used in single-event inference is
\begin{equation}
    \mathcal{L}(d|\theta) =\prod_{i}^{N_\text{det}} \prod^{M}_{j} \frac{1}{2 \pi S_{ij}} \exp{\left(-\frac{2}{T} \frac{| \tilde{d}_{ij} - \tilde{h}_{ij}(\theta) |^2}{S_{ij}} \right)} ,
    \label{eq:cbclike}
\end{equation}
where $d$ is the detector strain data, $\theta$ are the binary parameters, $\tilde{h}$ is the template for the response of the detector to the gravitational-wave signal as described in Section~\ref{sec:wf}, $S$ is the strain power spectral density, the products over $i$ and $j$ are over the $M$ frequency bins and $N_\text{det}$ detectors in the network respectively.

Different signals have different durations, and thus, different values for the number of frequency bins $M$.
The more frequency bins, the more embarrassingly parallel calculations to perform, and the more we expect to gain from the use of GPUs.
For example, systems with lower masses produce longer duration signals than systems with higher masses, all else equal.
In previously published observational papers, the minimum frequency is typically set to $\unit[20]{Hz}$ for binary black holes, although a larger minimum frequency was used in the analysis of GW170817~\cite{GW170817}.
For high mass binary black holes, this corresponds to $M=4{,}000-32{,}000$ bins.
Binary neutron star signals are much longer in duration, requiring $M\approx10^6$ bins.
When current detectors reach design sensitivity, frequencies down to $\unit[10]{Hz}$ will be used, leading to substantially longer signals and therefore many times more bins.
Future detectors are expected to be sensitive to frequencies as low as $\unit[5]{Hz}$~\cite{5Hz}.


During each likelihood evaluation, the most expensive step is usually calculating the template.
The next most expensive operation is applying a time translation to the frequency domain template in order to align the template with the merger signal.
To do this, the frequency-domain waveform is multiplied by a factor of $\exp{(-2i\pi f \delta t_{\text{det}})}$ where $t_\text{det}$ is different for each detector in the network.
This becomes increasingly expensive (with linear scaling) as more detectors are added to the network.

In Fig.~\ref{fig:cbclikespeedup} we show the speedup achieved evaluating the single-event likelihood function using our {\sc TaylorF2} implementation compared to the {\sc lalsimulation} implementation.
We find a speedup of an order of magnitude for a $\unit[128]{s}$-long signal, the typical duration for binary neutron star analysis with a minimum frequency of $\unit[20]{Hz}$.
This reduces the total calculation time from a few weeks to a few days.

When the number of frequency bins is relatively small $M<10^5$, we see a slowdown rather than a speedup.
This is due to the same overheads as seen in Section~\ref{sec:wf}.
Using the CUDA implementation of {\sc IMRPhenomPv2}, we would expect to see an acceleration for much shorter signal durations.

In order to demonstrate the improvement achieved
analyzing longer-duration signals, we analyze a synthetic binary neutron star inspiral two different ways.
First we analyze the final $\unit[128]{s}$ of inspiral from a frequency of $\unit[30]{Hz}$ as was done for the LIGO/Virgo analysis of GW170817~\cite{GW170817}.
Then we repeat the same analysis including $\unit[512]{s}$ worth of inspiral with a minimum frequency of $\unit[15]{Hz}$.

\begin{figure}[!t]
    \centering
    \includegraphics[width=\linewidth]{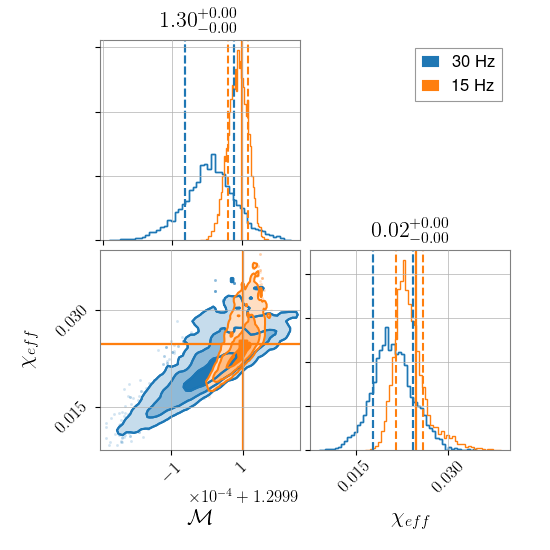}
    \caption{
    Posterior distributions for the chirp mass and effective spin of a binary neutron star inspiral similar to GW170817 when beginning the analysis at 30 Hz (blue) and 15 Hz (orange).
    Analyzing more of the early inspiral enables better measurement of the chirp mass, which leads to an improved measurement of the neutron star spins.
    }
    \label{fig:bns_compare}
\end{figure}

The one- and two-dimensional posterior distributions for chirp mass and effective aligned spin $\chi_\text{eff}$ are shown in Fig.~\ref{fig:bns_compare}.
Observing more of the early inspiral improves our measurement of both of these parameters.
A factor of $\sim 2$ improvement of the effective spin will facilitate future comparisons with the galactic pulsar population~\cite{Zhu2018}.

\section{\label{sec:popinf}Population acceleration}
In population inference, we are interested in measuring hyper-parameters, $\Lambda$, describing a \textit{population} of binaries (e.g., minimum/maximum black hole mass) rather than the parameters, $\theta$, of each of the individual binaries.
The population properties are often described by either phenomenological models (e.g.,~\cite{Vitale2017a,Talbot2017,Fishbach2017b,Kovetz2017a,Wysocki2017,Talbot2018a,Fishbach2018,Wysocki2018b,Roulet2019,Gaebel2019}) or by the results of detailed physical simulations, e.g., population synthesis or N-body dynamical simulations (e.g.,~\cite{Mandel2010,Stevenson2015,Belczynski2016b,Gerosa2017,Fishbach2017a,Stevenson2017b,Zaldarriaga2018,Zevin2017,Wysocki2018a,Barrett2018,Qin2018}).
In this work, we use the former for examples.
However, our methods apply equally to both.
The formalism for hierarchical inference including a discussion of selection effects is briefly described below
See, e.g.,~\cite{Farr2015, Thrane2019, Mandel2018} for detailed derivations.

In order to analyze a population of binary black holes, we typically use the following likelihood~\cite{Farr2015},
\begin{equation}
    \begin{split}
        \mathcal{L_{\text{tot}}}(\{d_{i}\} | \Lambda, R) &=  R^{N} e^{-RVT(\Lambda)} \prod^{N}_{i} \int d\theta_{i} \mathcal{L}({d_{i}} | \theta_{i}) p(\theta_{i} | \Lambda).
    \end{split}
\end{equation}
Here, $\mathcal{L}(d_{i} | \theta)$ is the likelihood of obtaining strain data $d_{i}$ given binary parameters $\theta_{i}$ as in Eq.~\ref{eq:cbclike}, $p(\theta|\Lambda)$ is our population model, and $VT(\Lambda)$ is the total observed spacetime volume if the population is described by $\Lambda$.
See, e.g.,~\cite{Finn1993, Dominik2015, Wysocki2018c, Tiwari2018} for discussions of methods to calculate $VT(\Lambda)$.

Within \textsc{GWPopulation} we currently support the calculation of $VT$ on a regular grid with GPU acceleration.
The calculation of $VT$ does not depend on the number of events.
However, this grid-based integration is limited to small dimensional spaces, and so the subdominant effects of spin on detectability cannot be included in this method.
This is mitigated by performing monte carlo integration.
However, the cost of this compation scales as $O(N)$~\cite{Farr2019}.
We are currently developing a method which will enable spin-effects to be included in the calculation of $VT$ wih no increase in cost at runtime~\cite{ML_VT}.
For the benchmarking performed in this work, we ignore this quantity, but it may be added back in without significant impact on performance.

Since $\mathcal{L}({d_{i}} | \theta)$ is independent of the population model, it can be evaluated independently for each observed binary, as described in Section~\ref{sec:cbc}.
Since we are interested in evaluating an integral of the likelihood over the binary parameters, it is convenient to perform a Monte Carlo integral using samples from the likelihood.
This process is known as ``recycling''.

The Bayesian inference algorithms used for single-event inference typically generate samples from the posterior distribution.
Therefore, it is necessary to weight each of the samples by the prior probability distribution $p(\theta_{ij} | \O)$ used during the initial inference step.
This yields the following likelihood
\begin{equation}
    \mathcal{L}_{\text{tot}}(\{d_i\}|\Lambda, R) \propto R^{N} e^{-RVT(\Lambda)} \prod^{N}_{i} \sum^{n_{i}}_{j} \frac{p(\theta_{ij}|\Lambda)}{p(\theta_{ij}|\O)}.
    \label{eq:poplike}
\end{equation}
Where $\{\theta_{j}\}_{i}$ is a set of $n_{i}$ samples drawn from the posterior distribution $p(\theta_{i} | d_{i})$.
The evaluation of the population model for each of the posterior samples is embarrassingly parallel.

\begin{figure}[!t]
    \centering
    \includegraphics[width=\linewidth]{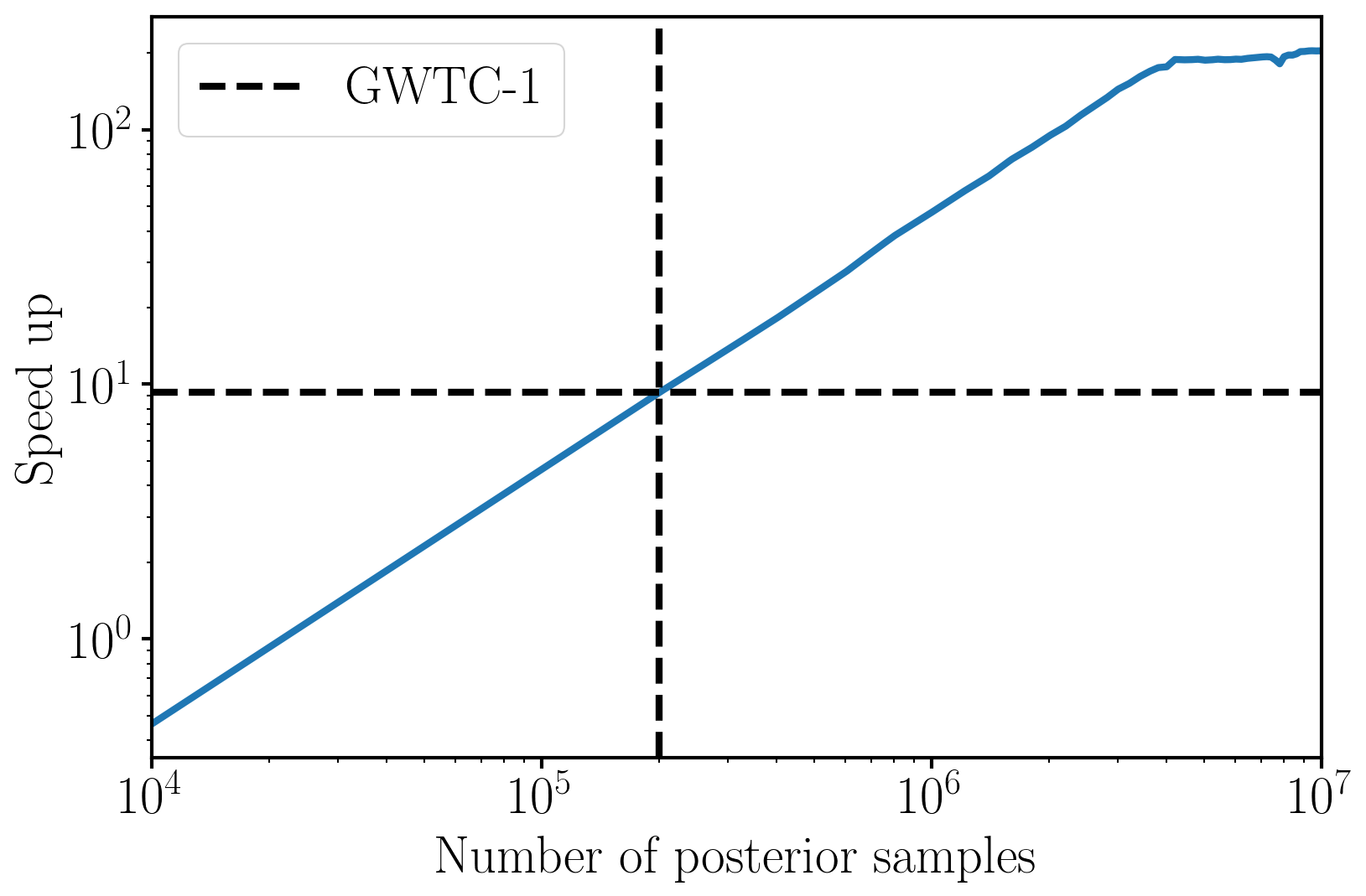}
    \caption{
    Ratio of the likelihood evaluation time (top) and likelihood evaluation time (bottom) on CPU and GPU as a function of number of samples.
    The vertical lines indicate (left to right): the number of samples released as part of the GWTC-1 data release, the anticipated number of samples at design sensitivity, the number of samples for a week of data.
    With the number of samples available as part of GWTC-1, the speedup is \gwtcspeedup.
    When the number of samples exceeds 4 million we reach the limits of the available GPUs and the likelihood evaluation time begins to increase.
    As GPU technology improves we expect that the maximum speedup over single-threaded code will continue to increase.
    }
    \label{fig:scaling}
\end{figure}

The first step is to draw an equal number of samples from each posterior so that for each binary parameter we have a single $N \times n_{i}$ array.
These arrays are then transferred to the GPU to evaluate the probabilities, sums, and logarithms.
We then need to only transfer a single number, the (log-)likelihood, back to the CPU when the likelihood is evaluated.



We calculate the likelihood evaluation time as a function of the number of posterior samples being recycled.
The tests performed in this work use the mass distribution proposed in~\cite{Talbot2018a}, the spin magnitude distribution from~\cite{Wysocki2018b}, and the spin orientation model from~\cite{Talbot2017}.
Fig.~\ref{fig:scaling} shows the expected linear scaling in the speedup obtained by using the GPU.
At around $3 \times 10^6$ samples, the GPU likelihood evaluation time begins increasing and the growth of the relative speedup slows.
This is due to GPU queue saturation.

The data released after the second observing run of advanced LIGO/Virgo includes ten binary black hole systems and the shortest posterior contains $\sim 2\times10^4$ posterior samples.
Using $2\times10^5$ samples the GPU code is $\approx10\times$ faster, reducing runtimes from a week to less than a day.

During Advanced LIGO/Virgo's third observing run, beginning April 2019, we can expect to detect tens more binary black hole mergers~\cite{Abbott2016}.
Given the current performance, we would expect the relative speed of the GPU code and the CPU code to continue to scale linearly with the size of the observed population.
Within the lifetime of current detectors, we can conservatively assume that we will detect hundreds of events.
At this stage using a GPU will accelerate population inference by more than two orders of magnitude.


\section{\label{sec:discussion}Discussion}
As the field of gravitational-wave astronomy grows, the quantity of data to be analyzed is rapidly increasing.
Thus, it is necessary to constantly improve and accelerate inference algorithms.
In this paper, we demonstrate multiple ways in which GPUs can aid in this endeavor.
We show that multiple orders of magnitude speedup can be achieved within the lifetime of current detectors in three areas:
\begin{itemize}
    \item waveform evaluation.
    \item CBC likelihood evaluation.
    \item population inference.
\end{itemize}
Most of these improvements use \textsc{cupy}, a python interface to \textsc{CUDA}, which acts as a GPU wrapper for existing C code.
\textsc{cupy} has also recently been used for other parameter estimation methods in gravitational-wave astronomy~\cite{Wysocki2019}.

We provide two complementary GPU versions of commonly used waveforms, a \textsc{CUDA} implementation of \textsc{IMRPhenomPv2} and a python implementation of \textsc{TaylorF2}.
We find that the performance of the \textsc{CUDA} waveform exceeds that of the pure-python waveform for short waveforms when efficient memory allocation is vital.
For longer waveforms, memory allocation is less important and the python waveforms give similar or greater speedups than the \textsc{CUDA} implementation.
The \textsc{CUDA} implementation of \textsc{IMRPhenomPv2} is available at~\footnote{\href{https://adacs-ss18a-rsmith-python.readthedocs.io/en/latest/}{adacs-ss18a-rsmith-python.readthedocs.io/en/latest/}}.
Future development of parallelized waveforms may enable rapid evaluation of waveforms encoding more of the phenomenology of general relativity, e.g., higher-order modes~\cite{Blackman2017b}, gravitational-wave kicks~\cite{Gerosa2018a}, or gravitational-wave memory~\cite{Talbot2018b}.

Other than waveform evaluation, the dominant cost for the likelihood used in inference for compact binary coalescences are exponentials to perform frequency domain time-shifts.
This is another operation which drastically benefits from parallelization.
Using these two methods, we reduce the likelihood evaluation time for binary neutron star mergers by an order of magnitude at current sensitivity and more when current detectors reach their design sensitivity.
The code for performing GPU-accelerated single-event parameter estimation can be found at~\footnote{\href{https://github.com/ColmTalbot/gpucbc}{github.com/ColmTalbot/gpucbc}}.

Other methods for speeding up likelihood evaluation for long signals include reduced order quadrature methods~\cite{Smith2016} and relative binning~\cite{Cornish2010, Zackay2018}.
These methods rely on the waveform being sufficiently well described by a small set of unevenly sampled frequencies.
For binary neutron star mergers like GW170817, the signal from $\sim 30$ Hz to the merger can be described with only $\sim 10^3$ frequencies.
Additionally, these methods do not require computing any exponentials at run time.
GPU waveforms will have less of a benefit for these cases.
However, we may be able to accelerate parameter estimation by an additional factor of a few.
This will facilitate more rapid production of sky maps for electromagnetic observers following up on gravitational-wave events.

The computational cost of performing population inference increases linearly with the size of the observed population.
Using a GPU to perform the embarrassingly parallel likelihood evaluation we find an acceleration of \gwtcspeedup using the data in GWTC-1~\cite{GWTC1} compared to the CPU code.
We additionally find that the GPU implementation will outperform the CPU code by more than two orders of magnitude during the lifetime of current detectors.
We therefore present \textsc{GWPopulation}~\footnote{\href{https://github.com/ColmTalbot/gwpopulation}{github.com/ColmTalbot/gwpopulation}, \textsc{GWPopulation} is also available through \textsc{pypi}.}: a CPU/GPU agnostic framework for performing gravitational-wave population inference.
Both of these packages use the framework available within \textsc{Bilby}~\cite{Ashton2019}.

Within \textsc{GWPopulation}, we include CPU/GPU tools for performing population inference.
We provide:
\begin{itemize}
    \item implementations of many previously proposed binary black hole population models.
    \item the likelihoods commonly used in gravitational-wave population inference.
    \item methods for computing selection biases.
\end{itemize}

Using our GPU-enabled implementation of a binary neutron inspiral waveform we demonstrate that beginning the analysis at lower frequencies will improve our measurements of the intrinsic parameters of the system.
While these results present a tantalizing glimpse of the physics that will be enabled through GPU acceleration, more work is required to realize these gains.
\begin{enumerate}
    \item When analyzing signals for many minutes it will be necessary to include the effect of the Earth's rotation in our analysis.
    Including the effect of the Earth's rotation will improve sky localization since the movement of the detector allows triangulation from data taken at different times.
    While progress on this front has been made in recent years, e.g.,~\cite{Marsat2018, Liang2019}, a working implementation is not available at this time.
    \item Central to the likelihood we use (Eq.~\ref{eq:cbclike}) is an assumption of gaussianity and stationarity of the noise.
    These assumptions are not generally valid over the lengths of time considered in this paper.
\end{enumerate}
The improvements between current sensitivity and the projected design sensitivities of advanced LIGO/Virgo are largest at low frequencies, so additional upgrades may be required in order to achieve the improvements shown here with real data.

As we enter ``the data era'' of gravitational-wave astronomy, optimizing Bayesian inference codes become ever more important.
When the likelihood evaluation requires a large number of independent operations, GPUs can yield significant benefits.

\acknowledgements
CT, RS, and ET are supported by the Australian Research Council (ARC) CE170100004.
ET is supported by ARC FT150100281.
This work used the OzStar supercomputing cluster and the LIGO Data Grid.
This work was supported by a Software Support grant from Astronomy Data And Computing Services (ADACS).
This is LIGO document P1900101.

\bibliography{refs.bib}

\end{document}